\documentclass[aps,prc,twocolumn,showpacs,superscriptaddress,groupaddress,amsmath,amssymb]{revtex4-1}

\usepackage{graphicx}  
\usepackage{dcolumn}   
\usepackage{bm}        
\usepackage{verbatim}   
\usepackage{textcomp}

\usepackage{color}

\begin{document}

\title{Bayesian Inference of Nuclear-Matter Density from Proton Scattering}


\author{J. C.~Zamora}
\email[E-mail address: ]{zamora@frib.msu.edu}
\affiliation{Facility for Rare Isotope Beams, Michigan State University, East Lansing, Michigan 48824, USA}


\date{\today}

\begin{abstract}
\begin{description}
\item[Background] Proton elastic scattering at intermediate energy is widely employed as a tool for determining the matter radius of atomic nuclei. The sensitivity of the approach relies on high-resolution measurements at small scattering angles and low-momentum transfer. Under these conditions, the Glauber multiple scattering theory accurately describes the proton-nucleus elastic cross section.

\item[Purpose] Investigate the sensitivity of the Glauber multiple scattering theory to uncertainties associated with input parameters such as the nuclear-matter density distribution and nucleon-nucleon data.

\item[Method]   A joint Bayesian inference was performed using 12 angular distributions of elastic scattering at different energies on ${}^{58}$Ni, ${}^{90}$Zr, and ${}^{208}$Pb targets. A Metropolis-Hastings algorithm was implemented to make an uncertainty quantification analysis for the input parameters used in the Glauber multiple scattering theory.

\item[Results] The experimental cross sections were  fitted simultaneously using a joint Bayesian inference approach. Posterior probability density distributions of 42 input parameters were obtained from the analysis. A moderate correlation between the nuclear density parameters and the nucleon-nucleon cross sections was found. This correlation impacts the extraction of the nuclear-matter radius.

\item[Conclusions]  The present analysis provided a consistent method for extracting the nuclear-matter density distribution of ${}^{58}$Ni, ${}^{90}$Zr, and ${}^{208}$Pb from data across different incident energies. Due to the correlation of the nucleon-nucleon cross sections with the other input parameters, a constrained Bayesian inference using free nucleon-nucleon cross section data was performed. The nuclear-matter radii obtained from the analysis are in good agreement with multiple results reported in the literature.

\end{description}

\end{abstract}
\maketitle

\section{Introduction}

One of the most fundamental properties of the atomic nucleus is its size. The nuclear size is often determined using electromagnetic probes, which are mainly sensitive to the charge-density distribution of the nucleus. Over the past several years, electron elastic scattering experiments have been successfully employed for extracting the charge radius of a wide range of stable nuclei \cite{RevModPhys.28.214}. New experimental techniques, like laser spectroscopy, have expanded our understanding of charge radii for short-lived nuclei \cite{GarciaRuiz2016,Koszorus2021}.
In contrast,  nuclear-matter (protons and neutrons) radii are typically not well-known. Most of the information about nuclear matter has been derived via hadron scattering, as it is highly sensitive to the nuclear force. For instance, proton elastic scattering at intermediate energy (hundreds of MeV) is widely used to extract the nuclear-matter size and its radial distribution \cite{ALKHAZOV197889}. The probe has been used with novel techniques to investigate the nuclear matter of unstable nuclei \cite{ALKHAZOV2002269, vonSchmid2023}. \par

The majority of the available proton elastic scattering data at intermediate energies were measured during the 70s and early 80s in facilities such as Saclay \cite{BERTINI1973119}, Gatchina \cite{Alkhazov_77}, and Los Alamos \cite{PhysRevLett.40.1256}. These measurements required high-resolution spectrographs that allowed the separation of elastically scattered protons at very forward angles from the unreacted beam. Measurements with an angular resolution below 1~degree were achieved with these detection systems. \par

The Glauber multiple scattering theory \cite{glauber1, glauber2} provides a good description of experimental proton elastic scattering cross sections at intermediate energy. The theory requires the input of a nuclear-matter density distribution for the target nucleus, which can be fine-tuned to improve agreement with experimental data. Thus, the analysis enables a model-dependent approach for determining the nuclear-matter radius. However, the resultant nuclear-matter radii and neutron skins obtained from analysis of multiple experiments at various proton incident energies exhibit significant differences \cite{ALKHAZOV197889, PhysRevLett.40.1256}. These discrepancies can be attributed to our poor knowledge of the nucleon-nucleon (NN) scattering amplitudes used in the Glauber multiple scattering theory. \par

Uncertainties arising from NN scattering data may influence the description of proton-nucleus collisions, particularly at energy levels where in-medium effects can be significant \cite{PhysRevC.20.1857, HUSSEIN1991279}. Nevertheless, uncertainties of NN parameters are generally not included in calculations of Glauber multiple scattering theory. Quantifying the uncertainty of all input parameters is important for understanding their sensitivity to the estimated cross section.  The present work presents a Bayesian uncertainty quantification (UQ) of Glauber multiple scattering theory using proton-nucleus elastic scattering data. The approach provides a consistent method for fitting multiple angular distributions within the same framework and for quantifying the systematic uncertainties associated with the nuclear-matter radii obtained from the analysis.\par
The paper is structured as follows:  Section~\ref{theory} gives a brief overview of the theoretical models employed in the study, namely the Glauber multiple scattering theory and Bayesian inference.  Section~\ref{results} presents the results of the Bayesian inference with the respective posterior probability density distributions (PDFs) for the model parameters used in the analysis. Finally, Section~\ref{summary} presents the conclusions of this work.

\section{Theoretical Models \label{theory}}

\subsection{Glauber Multiple Scattering Theory}

The Glauber multiple scattering theory \cite{glauber1, glauber2} provides a powerful framework for describing proton-nucleus scattering at intermediate- and high-energy collisions. The theory relies on the eikonal wavefunction, which is an approximation valid for scattering at small angles and low-momentum transfer. The Glauber formalism treats the proton-nucleus scattering as a series of multiple interactions between the incoming proton and individual nucleons within the nucleus. The approximation greatly simplifies the calculation of the total scattering amplitude, which can be obtained from a fundamental NN profile function. In particular, the elastic scattering cross section is derived microscopically by using NN scattering data as well as a nuclear density parameterization.  A brief description of the model is presented below. \par

The elastic cross section is calculated from the scattering amplitude as
\begin{equation}
 \frac{d\sigma}{d\Omega} = \left| F_{\text{el}}(q) \right|^2,
\end{equation}
where momentum transfer, $q$, is associated with the wave number of the incident proton, $k$, and the scattering angle $\theta$ in the center-of-mass frame, expressed as $q=2k\sin(\theta/2)$. For numerical evaluation, it is convenient to split the scattering amplitude in  Coulomb and nuclear parts as
\begin{equation}
 F_{\text{el}}(q) = F_{\text{C}}(q) + F_{\text{N}}(q).
\end{equation}
The Coulomb scattering amplitude can be calculated analytically by using the expression \cite{glauber2, bertulani2019introduction}
\begin{equation}
 F_{\text{C}}(q) = -\frac{2\eta k}{q^2} \exp{\left[ -2\eta i \ln{\frac{1}{2}q} + 2i\phi \right]},
\end{equation}
with $\eta = Ze^2/\hbar v$, $Z$ is the charge of the target nucleus, $v$ is the relative velocity between the incident proton and the nucleus, and $e$ is the elementary charge. The phase, $\phi$, is given by
\begin{equation}
 \phi = -\eta C + \sum_{w=0}^\infty \left(\frac{\eta}{w+1} -\arctan \left( \frac{\eta}{w+1} \right) \right),
\end{equation}
where $C=0.57721\dots$ is the Euler's constant. \par
The nuclear scattering amplitude is obtained by numerical integration of \cite{glauber2,ALKHAZOV1976443}
 \begin{align}
F_{\text{N}}(q) = -ik \int_0^\infty \left( e^{i\chi_\rho(b)} \left(1- \Gamma_p(b)\right)^Z\left(1- \Gamma_n(b)\right)^N  \right. \nonumber \\
\left. -e^{i\chi_0(b)} \right)\,b\,J_0(qb)\, \mathrm{d}b,
\end{align}
where $J_0$ is the zero-order Bessel function,  $\Gamma_{p(n)}$ is the proton (neutron) profile function, and $\chi_0$  and $\chi_\rho$  are Coulomb phase shifts. $\chi_0$ corresponds to the phase shift from a point-like charge distribution given by
\begin{equation}
 \chi_0(b) = 2\eta \ln(b).
\end{equation}
$\chi_\rho$ represents the Coulomb phase shift, which accounts for the finite charge distribution of the nucleus \cite{glauber2}
\begin{equation}
 \chi_\rho(b) = 4 \pi \eta \left[ \ln(b) \int_0^b T_c(b') b' \, \mathrm{d}b + \int_b^\infty T_c(b') \ln(b')b' \, \mathrm{d}b\right],
\end{equation}
where $T_c$ denotes the thickness function associated with the folding of proton and nuclear charge distributions as
\begin{equation}
 Tc(b)=\frac{1}{2\pi}\int_0^\infty S_p(q)\left(S_{cp}(q)\right)^2 q\,J_0(qb)\, \mathrm{d}q,
\end{equation}
with $S_{cp}$ and $S_p$ representing the proton and nuclear charge form factors, respectively. The nucleon profile function, $\Gamma_j$ ($j=p, n$), accounts for the phase shift of individual proton-nucleon collisions, and can be calculated as follows
\begin{equation}
 \Gamma_j(b) = -\frac{i}{k}\int_0^\infty S_j(q)f_j(q)q\,J_0(qb)\, \mathrm{d}q,
\end{equation}
where $f_j$ is the NN scattering amplitude that is commonly parameterized as \cite{glauber2,ALKHAZOV1976443, PhysRevC.20.1857,HUSSEIN1991279}
\begin{equation}
 f_j(q) = \frac{k \sigma_{pj}}{4\pi}\left(i+\alpha_{pj}\right) \exp \left(-\frac{q^2\beta^2_{pj}}{2}\right).
 \label{nn_par}
\end{equation}
Here, $\sigma_{pj}$ is the proton-nucleon cross section, $\alpha_{pj}$ is the ratio between the imaginary and the real part of the proton–nucleon scattering amplitude, and $\beta^2_{pj}$ is the slope parameter of the NN cross section. These parameters are extracted from experimental values of proton-proton and proton-neutron scattering and phase shift analysis \cite{PhysRevD.28.97, PhysRevD.54.1}. The spin-orbit term in the NN scattering amplitude was not considered in this work.  The spin-orbit effect contributes negligibly to scattering at low momentum transfer in intermediate- and high-energy collisions \cite{PhysRevA.23.1188}. Therefore, the inclusion of spin-orbit interaction has a minimal impact on the derived nuclear matter densities \cite{ALKHAZOV1980364}. \par

The nuclear form factor, $S_j$, is determined from the Fourier transform of the target nucleus density
\begin{equation}
 S_j(q) = 4 \pi \int_0^\infty \rho_j(r)\frac{\sin(qr)}{q}r\,\mathrm{d}r.
\end{equation}
The proton and neutron density distributions are considered to have the same shape ($\rho_n = (N/Z)\rho_p$). For the present analysis, a two-parameter Fermi function was assumed for the nuclear-matter density distribution, which is given by
\begin{equation}
 \rho(r) =  \frac{\rho_0}{1+\exp \left(\frac{r-R}{a} \right)},
 \label{den_par}
\end{equation}
where $\rho_0$ is a normalization factor, and $R$ and $a$ are the radius and diffuseness parameters, respectively.\par

A computer program written in C\texttt{++} was developed for calculating proton-nucleus elastic cross section using the  Glauber multiple scattering theory described above. Input parameters from Eqs.~(\ref{nn_par}) and (\ref{den_par}) are used to fit the experimental angular distributions. The NN scattering parameters are typically fixed for a specific incident proton energy, while the nuclear density parameters are optimized to describe the proton-nucleus elastic scattering data. The uncertainty associated with NN data is typically disregarded in most cases. The purpose of this study is to investigate the impact of uncertainties in all input parameters used in the Glauber model to calculate elastic scattering cross sections. A good alternative is to perform an UQ analysis with Bayesian statistics, as described in the following section.

\subsection{Bayesian Inference}
Bayesian inference establishes a consistent framework for quantifying uncertainties by integrating experimental data and PDFs derived from a certain model. The method entails the transition from a prior to a posterior probabilistic distribution, representing the learning process of incorporating new evidence or information into the analysis \cite{RevModPhys.83.943}. The prior and posterior PDFs denote the state of knowledge or level of uncertainty before and after the integration of experimental data. The connection between these probability distributions is formally defined in Bayes' rule as \cite{bernardo2009bayesian}:

\begin{equation}
    P(\theta \mid D) = \frac{L( \theta) P(\theta)}{Z}.
\end{equation}
The equation above represents the posterior PDF, denoted as $P(\theta \mid D)$, which is the conditional probability of the model parameters, $\theta$,  given the data, $D$. It is calculated from the product of  the prior distribution, $P(\theta)$, and the  likelihood function, $L(\theta)$, with a normalization factor \cite{bernardo2009bayesian}
\begin{equation}
  Z=  \int L(\theta) P(\theta) d\theta.
\end{equation}
The likelihood function provides information about the hypothesis test by varying $\theta$ while considering the observed data $D$ as fixed \cite{like}. Given a set of $n$ independent observations, $X = \{x_1, x_2, \ldots, x_n\}$, the joint likelihood function can be expressed as
\begin{equation} L(\theta) = \prod_{i=1}^n f(x_i \mid \theta),
\end{equation}
where $f$ denotes the probability mass function associated with the observation $x_i$. In this study, each $x_i$ represents an angular distribution, and the corresponding joint likelihood function is calculated as the product of the independent likelihood functions. \par

The normalization $Z$, also known as the marginal likelihood, is a constant that depends exclusively on the dataset and is independent of the parameters. However, the marginal likelihood generally does not have an analytical solution, posing a considerable constraint to Bayesian inference.  An alternative approach involves employing a Markov chain Monte Carlo (MCMC)  \cite{MCMC_rev} method to randomly sample from the unnormalized posterior distribution, which avoids explicitly calculating the marginal likelihood.  The main idea of the technique is to generate a sequence (chain) of sample parameters that converge to the posterior PDF. The sampling procedure employed in this study is the Metropolis-Hastings (MH) algorithm \cite{MH_method}, which is widely used in statistics and numerical simulations. The process involves generating sample parameters by performing a random walk towards a region in the parameter space that has a high probability. Thus, as the number of sample values increases, the distribution becomes a more accurate representation of the posterior PDF.\par

In the particular case of Bayesian inference  applied to Glauber multiple scattering theory, the model parameters ($\theta$) comprise the NN scattering data (Eq.~(\ref{nn_par})) and the nuclear density parameters (Eq.~(\ref{den_par})). The prior PDFs are assumed to follow normal distributions centered on their expected values, $\mu$, with a standard deviation of 10\% of the expected value ($\sigma=0.1\mu$). Different values of $\sigma$ were examined, yet no significant difference was observed in the posterior PDFs. This option represents an informative prior, enabling a more accurate inference and a deeper understanding of the connection between prior knowledge and the experimental data. A computer code written in C\texttt{++} was developed for performing the Bayesian inference. The code was fully coupled with the Glauber code described in the previous section.
A consistent UQ analysis using multiple proton-nucleus angular distributions from the literature was carried out through joint Bayesian inference. The results are presented in the next section.

\section{Results \label{results}}
Proton  elastic scattering cross sections on ${}^{58}$Ni \cite{PhysRevC.82.044611, PhysRevC.37.692, LOMBARD1981233}, ${}^{90}$Zr \cite{SAKAGUCHI20171,PhysRevLett.47.1436, PhysRevLett.40.1256, Alkhazov_77} and ${}^{208}$Pb \cite{PhysRevC.82.044611, HUTCHEON1988429, PhysRevC.21.1488, BERTINI1973119} at energies of 300, 500, 800, and 1050~MeV were employed for the joint Bayesian inference presented in this work.  The target nuclei were selected based on data available in the literature for the present incident energies.  In total, 12 angular distributions were fit simultaneously using Glauber multiple scattering theory. The model parameters for each incident proton energy consisted of NN cross sections ($\sigma_{pp}$ and $\sigma_{pn}$),  the ratio of the imaginary and real parts of the NN scattering amplitudes ($\alpha_{pp}$ and $\alpha_{pn}$), and the slope parameter of the NN cross sections ($\beta^2_{pp}$ and $\beta^2_{pn}$). The prior PDF centroid values for $\sigma_{pp}$ and $\sigma_{pn}$ were derived from a least squares fitting of experimental data presented in Ref.~\cite{PhysRevC.81.064603}, while the values for $\alpha_{pp}$, $\alpha_{pn}$, $\beta^2_{pp}$, and $\beta^2_{pn}$ were obtained from the systematic study of proton-nucleus scattering reported in Ref.~\cite{PhysRevC.20.1857}. The nuclear density parameters ($R$ and $a$) were considered as model parameters specific to each target nucleus. The centroid values for the corresponding prior PDFs were based on density parameters extracted from electron scattering experiments \cite{DeVries1987495}. Finally, scaling factors for the theoretical cross sections were incorporated in the model to match the experimental data. The prior PDFs of these parameters were assumed to be distributions centered around 1. A total of 42 parameters were evaluated for the Bayesian inference using 12 angular distributions corresponding to different incident proton energies and target nuclei. In order to ensure convergence and to improve the accuracy of the posterior distribution estimates, the Bayesian inference was split into 448 parallel Markov chains, each comprising $2\times10^4$ samples. The random walk steps were dynamically adjusted during the sampling process to improve efficiency and convergence of the MH algorithm. \par
\begin{figure*}[!ht]
\centering
\includegraphics[width=1.0\textwidth]{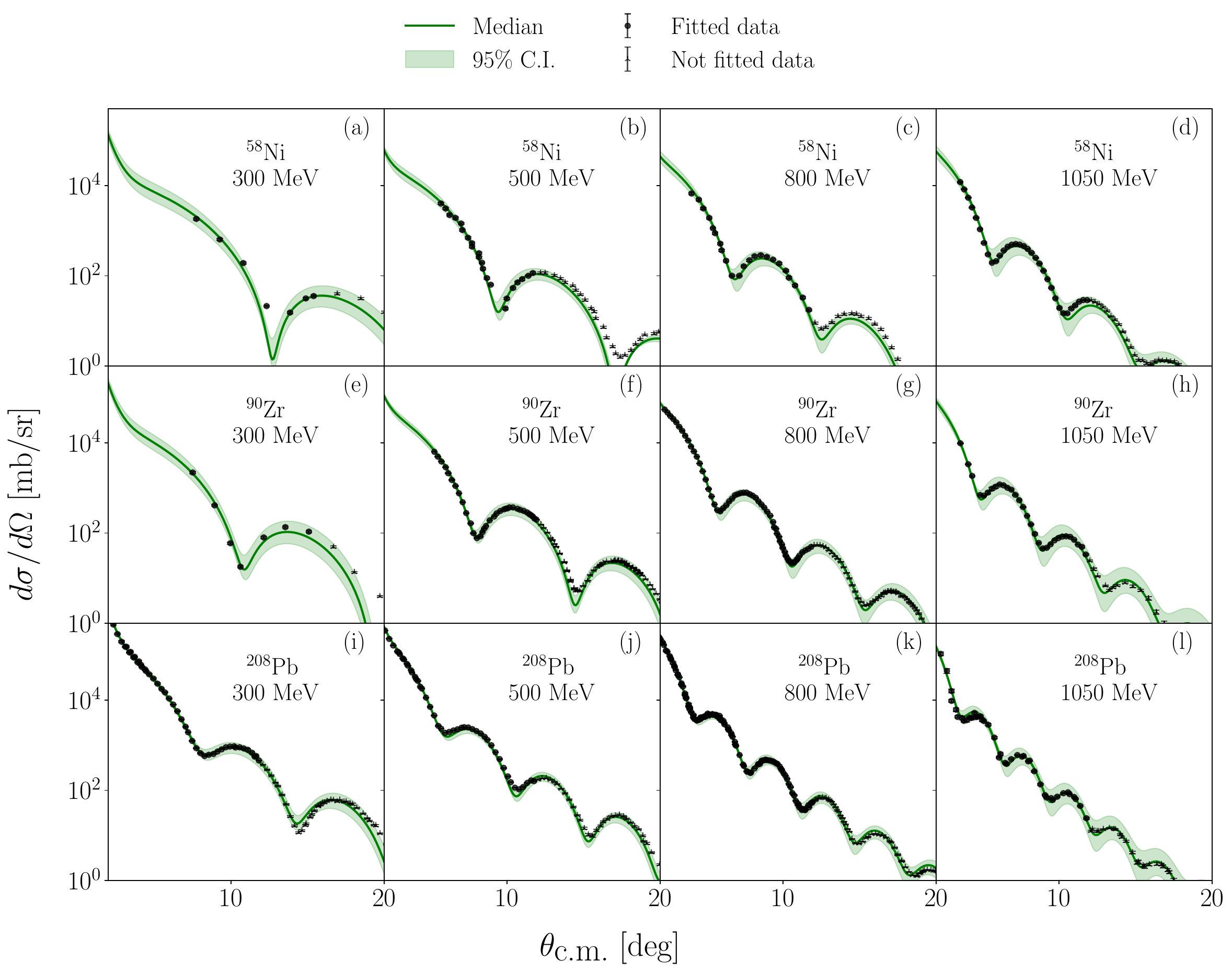}
\caption{\label{adist}  Angular distributions of proton elastic scattering on ${}^{58}$Ni (a, b, c , d), ${}^{90}$Zr (e, f, g , h) and ${}^{208}$Pb (i, j, k , l) at selected incident energies. The points are data obtained from multiple experiments \cite{PhysRevC.82.044611, PhysRevC.37.692, LOMBARD1981233, SAKAGUCHI20171,PhysRevLett.47.1436, PhysRevLett.40.1256, Alkhazov_77,PhysRevC.82.044611, HUTCHEON1988429, PhysRevC.21.1488, BERTINI1973119}. Only data up to 15 degrees were included for the Bayesian inference analysis. The solid lines and shaded bands represent the median values and uncertainties derived from the posterior PDFs.}
\end{figure*}

The resulting angular distributions from the Bayesian inference are shown in Figure~\ref{adist}.
The shaded bands denote the 95\% credibility interval (C.I.), while the solid lines indicate the median of the angular distributions. The points represent experimental data that are divided into fitted and not fitted data. Only data below 15 degrees were considered for Bayesian inference, as the Glauber scattering theory is valid for small scattering angles and low momentum transfer. This angular threshold was chosen to cover the region just above the first minimum of the cross section for ${}^{58}$Ni at 300 MeV, which is the case with less forward kinematics. The same threshold was used for consistency with all the other angular distributions. Multiple angular thresholds ranging from 10 to 20 degrees  were tested with no significant implications for the results.  As can be noticed, the model is able to describe the experimental data even for angles above 15 degrees, which were not included in the fit. As previously described, NN parameters were common for ${}^{58}$Ni, ${}^{90}$Zr, and ${}^{208}$Pb  data at a given incident energy, whereas density parameters were the same for a specific nucleus. In other words, the density distributions extracted from the analysis are consistent across the different incident energies studied in this work.  The respective uncertainty in the angular distributions arises from the propagation of the posterior PDF of each model parameter.

\begin{figure*}[!ht]
\centering
\includegraphics[width=1.0\textwidth]{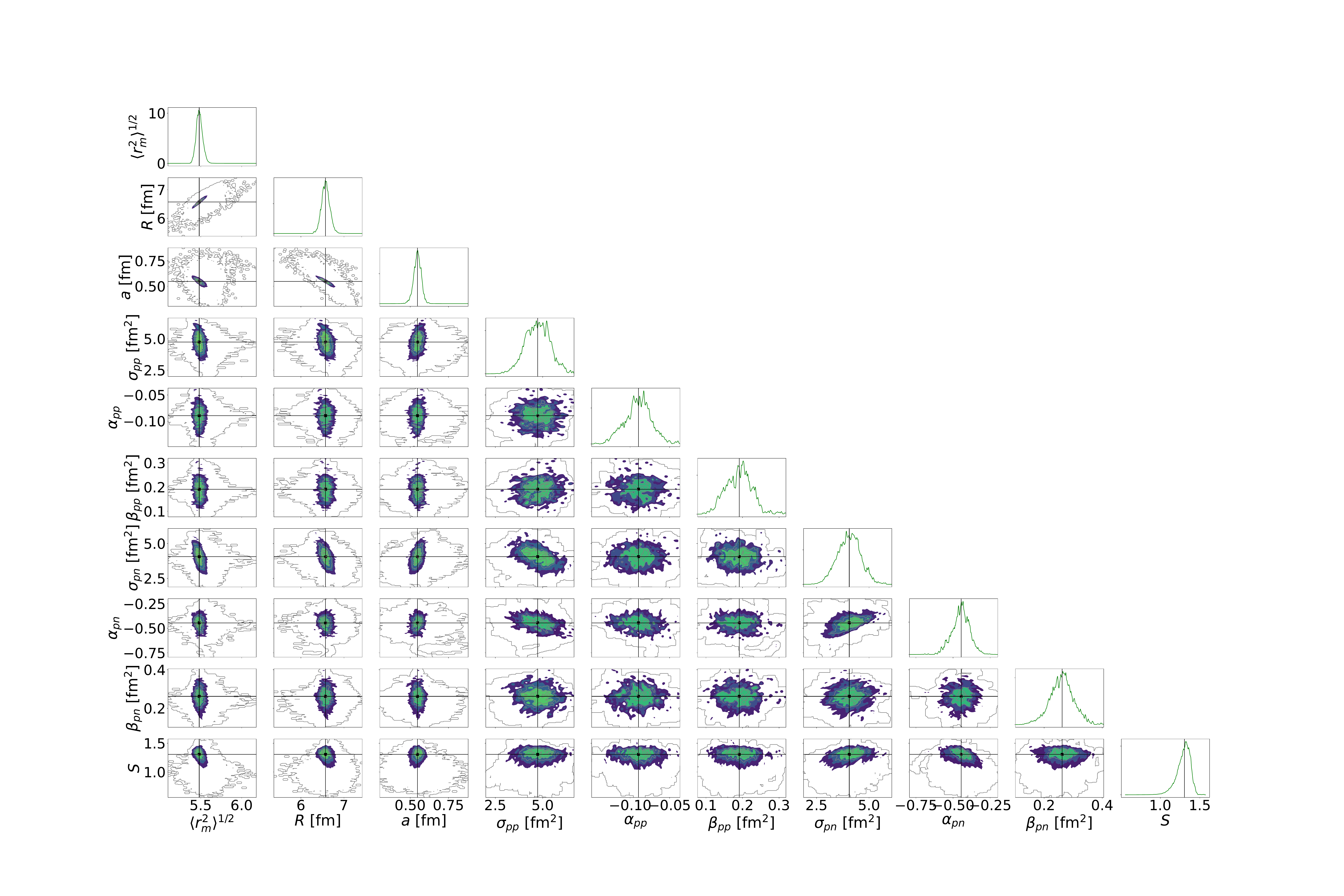}
\caption{\label{pair1}  Correlation plots of the posterior PDFs. The figure only shows a sample of the model parameters involved in the Bayesian inference. The parameters correspond to the nucleus ${}^{208}$Pb and the incident energy of 1050~MeV. The light green color in the plots represent the region with the highest probability.  }
\end{figure*}

Figure~\ref{pair1} displays a sample of the posterior PDF correlation plots that correspond to the incident energy of 1050~MeV and the density parameters for  ${}^{208}$Pb. The root-mean-square (RMS) radius, $\langle r_m\rangle^{1/2}$, was not included in the data fitting process, but it can be calculated using the following expression:

\begin{equation}
 \langle r_m^2\rangle^{1/2} = \left( \frac{\int_0^\infty r^4 \rho(r)\,\mathrm{d}r}{\int_0^\infty r^2 \rho(r)\,\mathrm{d}r} \right)^{1/2},
\end{equation}
where $\rho$ is the matter density distribution as defined in Eq.~(\ref{den_par}). The parameter $S$ in Figure~\ref{pair1} is a scaling factor employed to match the calculated angular distribution with the experimental data. A moderate correlation is observed for the density parameters in relation to all other posterior PDFs. In particular, the nuclear radius ($R$) and diffuseness ($a$) are correlated with the NN cross sections. The Pearson correlation coefficient of $R$ with respect to $\sigma_{pp}$ ($\sigma_{pn}$) is approximately -0.5, while the correlation coefficient between $a$ and $\sigma_{pp}$ ($\sigma_{pn}$) is roughly 0.4.  This means the values of $\sigma_{pp}$ and $\sigma_{pn}$ (and their  uncertainties) may affect the extraction of the nuclear-matter radius. Table~\ref{table1} presents the Bayesian inference results for all model parameters except the nuclear density values. The numbers shown represent the median of the corresponding posterior PDF, with the uncertainty indicated by the 95\% credibility interval. The parameters $S_1$, $S_2$, and $S_3$ are the cross-section scaling factors for  ${}^{58}$Ni, ${}^{90}$Zr, and ${}^{208}$Pb, respectively. The relative uncertainty (one standard deviation) for all parameters in this analysis varies between 10\% and 20\%. These uncertainties are slightly higher than the experimental NN cross sections \cite{ptac097, PhysRevD.11.529}, which range from 1\% to 9\% for the incident energies evaluated in the present study. However, the uncertainties $\alpha_{pp}$, $\alpha_{pn}$, $\beta^2_{pp}$, and $\beta^2_{pn}$ are consistent with experimental data from Refs.~\cite{ptac097, RevModPhys.50.523}, which reported values ranging from 10\% to 30\%.

\begin{table*}[!ht]
\renewcommand{\arraystretch}{1.5} 
\caption {\label{table1} Results of the Bayesian inference analysis. The values correspond to the median of the posterior PDFs and the uncertainties represent the 95\% credibility interval.}
\begin{ruledtabular}
\begin{tabular}{cccccccccc}
  Energy [MeV]   & $\sigma_{pp}$ [fm$^{2}$] & $\alpha_{pp}$ &  $\beta_{pp}$ [fm$^{2}$] & $\sigma_{pn}$ [fm$^{2}$] & $\alpha_{pn}$ &  $\beta_{pn}$ [fm$^{2}$] & $S_1$ & $S_2$ & $S_3$      \\   \hline
 300 & $2.46_{-0.68}^{+0.73}$
 & $0.64_{-0.19}^{+0.18}$ & $0.63_{-0.22}^{+0.21}$  & $3.47_{-0.66}^{+0.65}$   & $0.29_{-0.08}^{+0.09}$ & $0.83_{-0.28}^{+0.28}$ & $0.83_{-0.19}^{+0.23}$ & $1.04_{-0.24}^{+0.22}$ &  $0.93_{-0.05}^{+0.05}$ \\
   500 & $3.13_{-0.77}^{+0.86}$  & $0.35_{-0.11}^{+0.10}$ & $0.21_{-0.08}^{+0.09}$  & $4.16_{-0.81}^{+0.83}$  & $-0.04_{-0.01}^{+0.02}$ &$0.29_{-0.11}^{+0.11}$ & $0.87_{-0.12}^{+0.13}$  & $0.88_{-0.07}^{+0.06}$ & $0.86_{-0.03}^{+0.02}$  \\
   800 & $3.82_{-1.09}^{+1.31}$ & $0.05_{-0.02}^{+0.02}$ & $0.18_{-0.06}^{+0.08}$ & $5.15_{-1.06}^{+0.77}$  & $-0.30_{-0.06}^{+0.08}$ & $0.25_{-0.08}^{+0.08}$ & $1.13_{-0.20}^{+0.15}$  & $1.12_{-0.07}^{+0.05}$ & $0.99_{-0.06}^{+0.03}$ \\
   1050 & $5.07_{-1.37}^{+1.44}$  & $-0.09_{-0.04}^{+0.03}$ & $0.19_{-0.05}^{+0.06}$  & $3.99_{-1.00}^{+1.05}$   & $-0.45_{-0.12}^{+0.10}$ & $0.26_{-0.09}^{+0.08}$ & $0.93_{-0.14}^{+0.13}$  & $1.02_{-0.12}^{+0.10}$ & $1.27_{-0.16}^{+0.14}$ \\
\end{tabular}
\end{ruledtabular}

\end{table*}

\begin{figure}[!ht]
\centering
\includegraphics[width=0.5\textwidth]{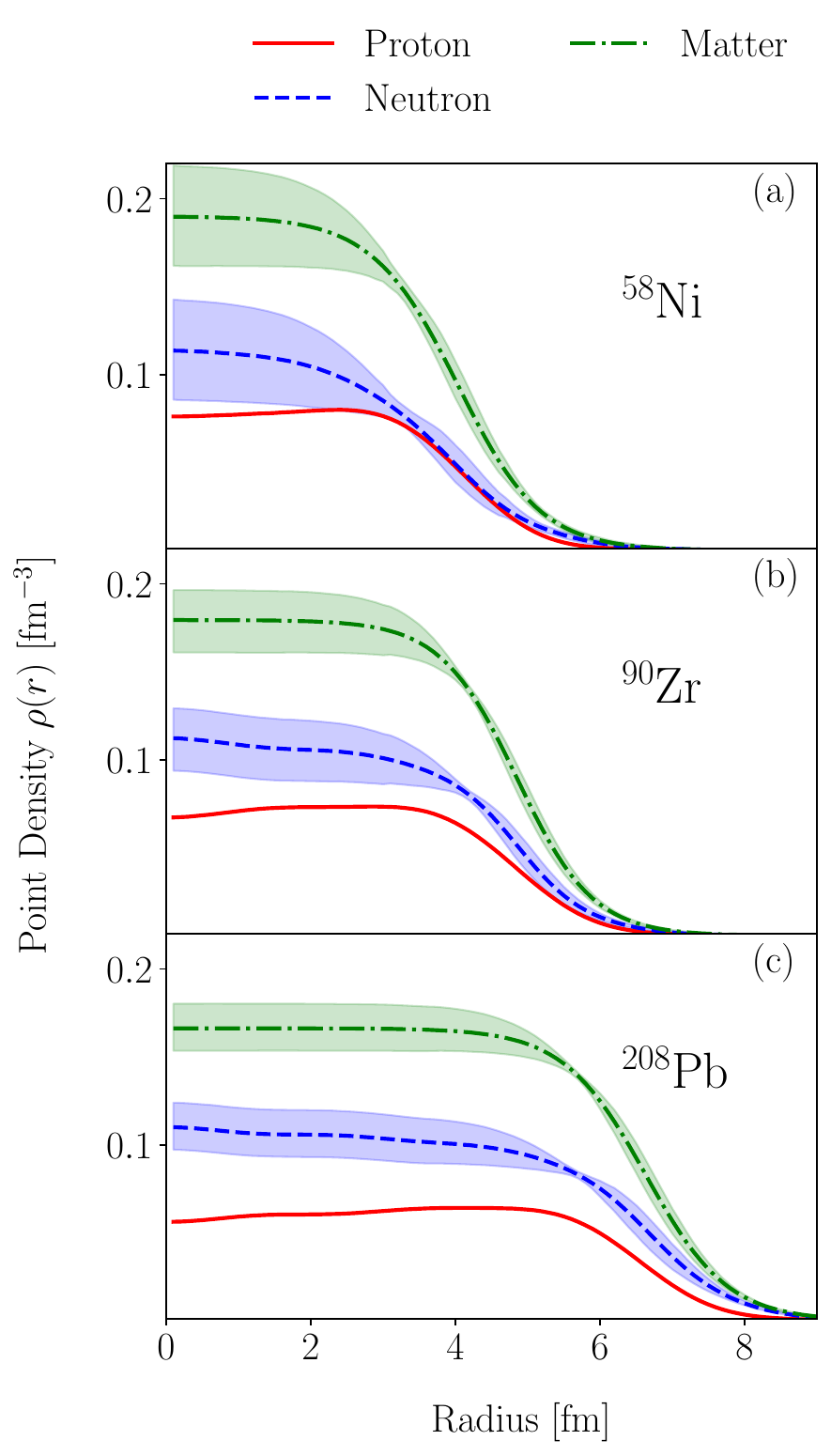}
\caption{\label{Ndens}  Point density distributions for for ${}^{58}$Ni (a), ${}^{90}$Zr (b), and ${}^{208}$Pb (c) derived from the Bayesian inference analysis. The lines correspond to the median value of the posterior PDF, while the shaded bands represent the 95\% credibility interval. The proton density distributions were obtained from electron scattering measurements.  }
\end{figure}

Figure~\ref{Ndens} shows the point nuclear-matter densities obtained from the Bayesian inference. The respective density parameters and RMS radii are presented in Table~\ref{table2}. Proton densities were derived from charge density parameterizations based on electron scattering measurements \cite{DEJAGER1974479,FAJARDO1971363,PhysRevLett.23.1402}.  The finite size of the nucleons  was unfolded from the charge density distributions using a Fourier transform with the phenomenological parameterizations for the nucleon form factors from Refs.~\cite{PhysRevC.66.065203, PhysRevC.70.068202}. It is important to note that more recent parameterizations, such as those presented in Refs.~\cite{PhysRevC.75.035202,PhysRevD.102.074012}, yield a result for the proton radius 3\% smaller than the one used in this work. However, the effect on the unfolded proton and neutron radii is only about 0.1\%. The resulting RMS proton radii from the unfolded distributions are 3.68, 4.20, and 5.44~fm for ${}^{58}$Ni, ${}^{90}$Zr, and ${}^{208}$Pb, respectively.  The neutron densities were obtained by subtracting the proton density from the nuclear-matter density. The shaded bands in Figure~\ref{Ndens} represent the 95\% credibility interval for both cases.  The extracted RMS nuclear-matter radius (see Table~\ref{table2}) of ${}^{58}$Ni is in agreement with other proton scattering analyses that reported values ranging from $3.65(5)$~fm \cite{Alkhazov1977402}  to $3.74(13)$~fm \cite{PhysRevC.100.054609}. The RMS nuclear-matter radius of ${}^{90}$Zr is consistent within the error bars with previously reported analyses, specifically $4.17(1)$~fm \cite{PhysRevC.108.054610}  and $4.23(1)$~fm \cite{HUANG2023138293}. The RMS nuclear-matter radius of ${}^{208}$Pb obtained from the inference is also consistent with other values obtained from proton scattering experiments, such as $5.50(5)$~fm \cite{ALKHAZOV1982430} and  $5.55(7)$~fm \cite{PhysRevC.18.1436}.

\begin{table}[!ht]
\renewcommand{\arraystretch}{1.5} 
\caption {\label{table2} Density parameters obtained from the Bayesian inference.  The values correspond to the median of the posterior PDFs and the uncertainties represent the 95\% credibility interval. }
\begin{ruledtabular}
\begin{tabular}{ccccc}
     & $R$ & $a$ &  $\langle r^2_m\rangle^{1/2}$ & $\langle r^2_n\rangle^{1/2}$       \\   \hline
 ${}^{58}$Ni &$3.92_{-0.30}^{+0.29}$ &$0.57_{-0.10}^{+0.08}$ & $3.71_{-0.07}^{+0.07}$ & $3.74_{-0.07}^{+0.07}$   \\
 ${}^{90}$Zr &$4.75_{-0.19}^{+0.20}$ & $0.52_{-0.05}^{+0.04}$ & $4.16_{-0.08}^{+0.08}$ & $4.12_{-0.08}^{+0.08}$  \\
 ${}^{208}$Pb & $6.53_{-0.22}^{+0.20}$ & $0.56_{-0.05}^{+0.05}$ & $5.47_{-0.09}^{+0.08}$ & $5.50_{-0.09}^{+0.08}$  \\
\end{tabular}
\end{ruledtabular}
\end{table}

\begin{figure}[!ht]
\centering
\includegraphics[width=0.45\textwidth]{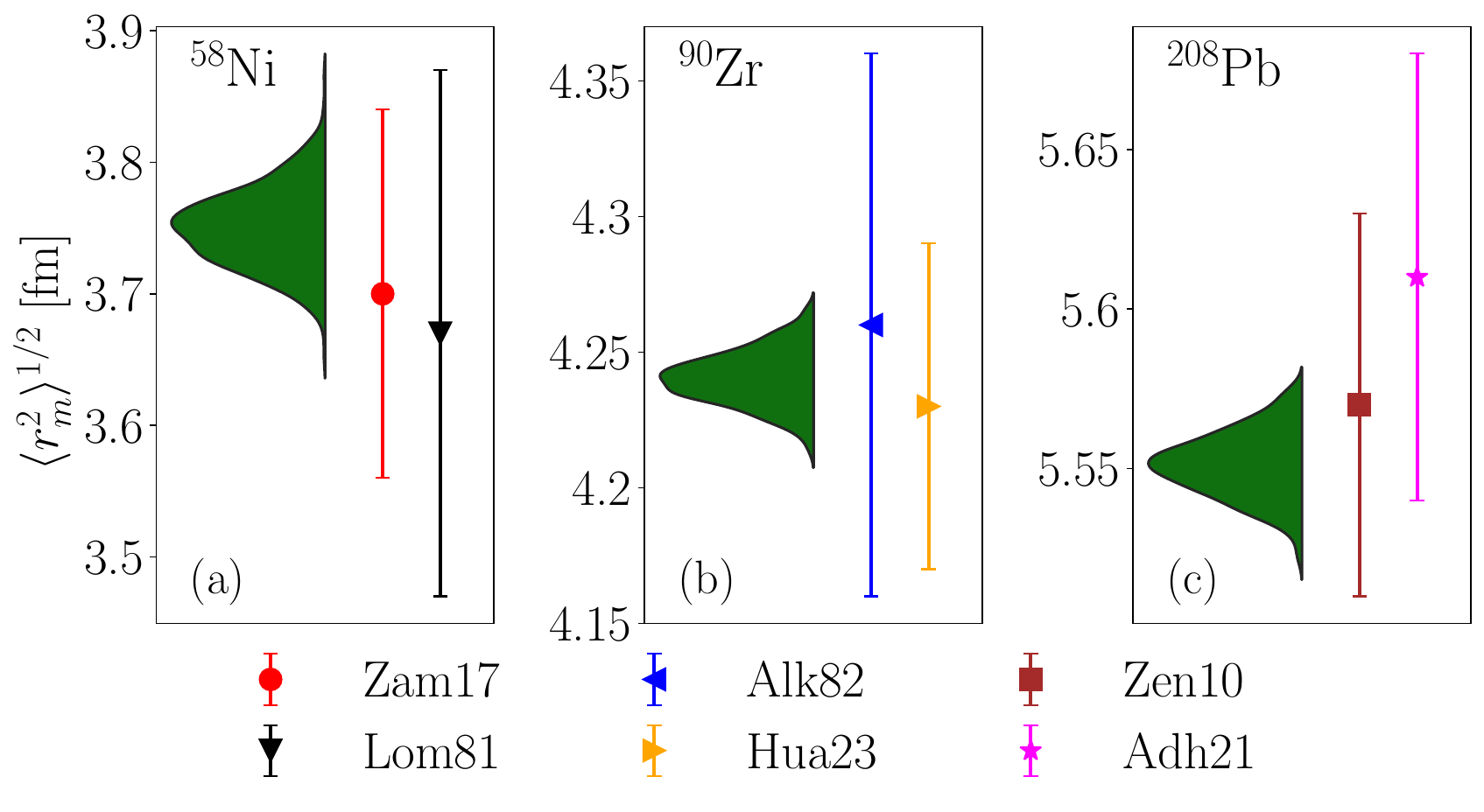}
\caption{\label{rms}  Probability distribution of the RMS nuclear-matter radius for ${}^{58}$Ni (a), ${}^{90}$Zr (b), and ${}^{208}$Pb (c). The distributions are compared with results from literature: Zam17 \cite{PhysRevC.96.034617}, Lom81 \cite{LOMBARD1981233}, Alk82 \cite{ALKHAZOV1982430}, Hua23 \cite{HUANG2023138293}, Zen10 \cite{PhysRevC.82.044611} and Adh21 \cite{PhysRevLett.126.172502}. For a better comparison with the distributions, the uncertainties were taken as $\pm 2\sigma$ for each data point.  }
\end{figure}

The values of $\sigma_{pp}$ and $\sigma_{pn}$ derived from the present analysis are comparable to the free NN cross-section data, but discrepancies can be up to 35\% from their expected values. The correlation between $\sigma_{pp}$ and $\sigma_{pn}$ with the nuclear density parameters can lead to an underestimation of the RMS radius. The Bayesian inference can be constrained by fixing $\sigma_{pp}$ and $\sigma_{pn}$ according with the parameterization presented in Ref.~\cite{PhysRevC.81.064603}. Due to this constraint, the number of model parameters used in the Bayesian inference analysis is reduced from 42 to 34. Table~\ref{table3} shows the Bayesian inference results using the constrained model. As can be noticed, there are no significant effect on the NN parameters or scaling factors after using the free NN cross sections. However, adding this prior information to the model has a strong impact on the density parameters (and, consequently, the RMS radii), as presented in Table~\ref{table4}. The median RMS radii increased systematically across all nuclei, while the associated uncertainties are significantly smaller than the results from the unconstrained model (Table~\ref{table2}). These new extracted values are fully consistent within the error bars with multiple results from literature \cite{Alkhazov1977402, PhysRevC.96.034617, LOMBARD1981233, PhysRevC.100.054609,ALKHAZOV1982430,HUANG2023138293, PhysRevC.108.054610,PhysRevC.82.044611,PhysRevC.18.1436, PhysRevLett.126.172502}.   Figure~\ref{rms} shows a comparison of the probability distribution of the RMS nuclear-matter radii with data from literature. In order to have a better comparison, the error bars of the data from literature correspond to $\pm 2\sigma$. The centroid of the RMS radius of ${}^{58}$Ni is above the reported values from Refs.~\cite{PhysRevC.96.034617,LOMBARD1981233}. Nevertheless, the results remain consistent when the distribution is compared with the corresponding error bars of the data.  In the case of ${}^{90}$Zr, the RMS radius is fully compatible with the other analyses. It is important to note that the uncertainty of the extracted RMS radius in this work is significantly smaller than the reported values. Finally, the RMS radius of ${}^{208}$Pb has attracted a renewed interest due to the recent results of the PREX measurement using parity-violating electron scattering \cite{PhysRevLett.126.172502}. A relatively large weak neutron skin, $\Delta r^w_{np}=0.28(7)$~fm,   was reported for ${}^{208}$Pb. Rather than comparing the extracted neutron skin, one may compare the nuclear-matter radius obtained from the PREX measurement with the well-established proton RMS radius (Figure~\ref{Ndens}). Figure~\ref{rms}(c) shows that the RMS radius derived from the PREX measurement remains large relative to the RMS distribution obtained in this analysis and other literature data. However, when comparing the uncertainties of the values, the RMS radius from this study agrees with the results of the PREX experiment.

\begin{table*}[!ht]
\renewcommand{\arraystretch}{1.5} 
\caption {\label{table3} Results of the constrained Bayesian inference analysis. The values correspond to the median of the posterior PDFs and the uncertainties represent the 95\% credibility interval. The parameters marked with $(*)$ were fixed during the analysis.}
\begin{ruledtabular}
\begin{tabular}{cccccccccc}
  Energy [MeV]   & $\sigma_{pp}^{(*)}$ [fm$^{2}$] & $\alpha_{pp}$ &  $\beta_{pp}$ [fm$^{2}$] & $\sigma_{pn}^{(*)}$ [fm$^{2}$] & $\alpha_{pn}$ &  $\beta_{pn}$ [fm$^{2}$] & $S_1$ & $S_2$ & $S_3$      \\   \hline
 300 & 2.33 & $0.63_{-0.18}^{+0.17}$
 & $0.62_{-0.23}^{+0.23}$  & 3.55  & $0.29_{-0.08}^{+0.08}$   & $0.78_{-0.30}^{+0.28}$  & $0.81_{-0.20}^{+0.27}$  & $1.03_{-0.23}^{+0.23}$ & $0.95_{-0.03}^{+0.03}$     \\
 500 & 3.07
 & $0.34_{-0.08}^{+0.08}$ & $0.22_{-0.08}^{+0.07}$  & 3.34     & $-0.04_{-0.02}^{+0.02}$ & $0.31_{-0.10}^{+0.10}$ & $0.82_{-0.11}^{+0.12}$   & $0.78_{-0.03}^{+0.0}$ & $0.83_{-0.01}^{+0.01}$  \\
 800 & 4.76
 & $0.05_{-0.02}^{+0.02}$  & $0.17_{-0.06}^{+0.06}$   & 3.78   & $-0.33_{-0.08}^{+0.10}$ & $0.26_{-0.08}^{+0.09}$  & $1.09_{-0.17}^{+0.13}$   & $1.07_{-0.05}^{+0.04}$  &  $0.95_{-0.04}^{+0.04}$ \\
 1050 & 4.74
 & $-0.09_{-0.03}^{+0.04}$ & $0.18_{-0.07}^{+0.07}$  &  3.86  & $-0.43_{-0.08}^{+0.09}$ &$0.25_{-0.08}^{+0.07}$  & $0.91_{-0.14}^{+0.14}$ & $0.99_{-0.08}^{+0.08}$ &  $1.26_{-0.14}^{+0.12}$ \\
\end{tabular}
\end{ruledtabular}
\end{table*}

\begin{table}[!ht]
\renewcommand{\arraystretch}{1.5} 
\caption {\label{table4} Density parameters obtained from the constrained Bayesian inference. The NN cross sections were fixed during the analysis (see Table~\ref{table3}).  The values correspond to the median of the posterior PDFs and the uncertainties represent the 95\% credibility interval.}
\begin{ruledtabular}
\begin{tabular}{ccccc}
     & $R$ & $a$ &  $\langle r^2_m\rangle^{1/2}$ & $\langle r^2_n\rangle^{1/2}$       \\   \hline
 ${}^{58}$Ni &$4.09_{-0.17}^{+0.15}$ & $0.54_{-0.09}^{+0.09}$ & $3.75_{-0.07}^{+0.07}$ & $3.81_{-0.07}^{+0.07}$ \\
 ${}^{90}$Zr &$4.90_{-0.03}^{+0.03}$ & $0.51_{-0.02}^{+0.02}$  & $4.24_{-0.02}^{+0.02}$ & $4.27_{-0.02}^{+0.02}$  \\
 ${}^{208}$Pb & $6.69_{-0.03}^{+0.02}$ &$0.53_{-0.02}^{+0.02}$ & $5.55_{-0.02}^{+0.02}$ & $5.62_{-0.02}^{+0.02}$  \\
\end{tabular}
\end{ruledtabular}
\end{table}

\section{Summary \label{summary}}
A Bayesian inference of nuclear-matter density distributions was performed using proton elastic scattering data and the Glauber multiple scattering theory. The analysis used 12 angular distributions at incident energies of 300, 500, 800, and 1050~MeV for ${}^{58}$Ni, ${}^{90}$Zr, and ${}^{208}$Pb target nuclei. The experimental cross sections were all fitted simultaneously using a joint Bayesian inference approach. The analysis provided a consistent method for extracting the nuclear-matter density distribution of ${}^{58}$Ni, ${}^{90}$Zr, and ${}^{208}$Pb from data across different incident energies.\par

Proton-nucleus elastic scattering cross sections were calculated from the Glauber multiple scattering theory that requires input from NN scattering data and a nuclear-matter density distribution.  The Bayesian inference implemented in the present work allows for consistent quantification of the uncertainties of all the input variables, particularly the NN parameters, which are frequently omitted. It was found that the uncertainties associated with the NN parameters have a significant impact on the extraction of the nuclear-matter density distribution. Specifically, the density parameters are highly sensitive to the NN cross sections. Therefore, including systematic uncertainties from NN data parameterizations into the Glauber model may improve the accuracy of the obtained nuclear-matter density distributions. \par

The uncertainties associated with the nuclear matter density are considerably reduced after imposing a constrain in the NN cross sections. The RMS nuclear-matter radii resulting from the analysis are in good agreement with multiple results reported in the literature.  In particular, the RMS nuclear matter of ${}^{208}$Pb was compared to the value obtained from the most recent parity-violating electron scattering measurement. The RMS radius that was obtained from the present analysis is consistent with the weak matter radius of ${}^{208}$Pb within the 95\% credibility interval.

\section*{Acknowledgements}
This work is dedicated to the memory of Juan Zamora (father). I am grateful to Professors Luiz Chamon (USP) and Adriana Barioni (UNIFESP) for their careful reading and comments on the manuscript. This material is based upon work supported by the U.S. Department of Energy, Office of Science, Office of Nuclear Physics and used resources of the Facility for Rare Isotope Beams (FRIB), which is a DOE Office of Science User Facility, operated by Michigan State University, under Award Number DE-SC0000661.

\bibliography{bibliography}  

\end{document}